\documentclass[useAMS,usenatbib,10pt,epsfig]{mn2e}
\usepackage[english]{babel}
\usepackage{amssymb}
\usepackage{graphicx}
\title[Rapid HeII$\rightarrow$HeI recombination]
{Rapid HeII$\rightarrow$HeI recombination  
\\and radiation concerned with this process}
\author[Kholupenko, Ivanchik, and Varshalovich]
{E.E. Kholupenko$^{1}$, 
A.V. Ivanchik$^{1,2}$, and 
D.A. Varshalovich$^{1,2}$\\
$^{1}$Ioffe Physical-Technical Institute, Saint-Petersburg 194021, Russia\\
$^{2}$Academical Physical-Technological University, Saint-Petersburg 194021, Russia}
\begin{document}

\date{01 January 2007}

\pagerange{\pageref{firstpage}--\pageref{lastpage}} \pubyear{2002}

\maketitle

\label{firstpage}
\begin{abstract}
Recombination of the primordial helium plasma (HeII$\rightarrow$HeI, 
$z\simeq 1500 - 3000$) is considered. This process has an effect on 
the CMBR anisotropy and CMBR spectrum 
distortion. In this work an influence of neutral hydrogen on kinetics of 
HeII$\rightarrow$HeI 
recombination is investigated in the frame of the standard cosmological model. 
It is shown that small amount of neutral hydrogen 
($10^{-5} - 10^{-2}$ of total number of hydrogen ions and atoms) 
leads to acceleration of HeII$\rightarrow$HeI recombination at 
$z\lesssim 2000$ and at $z\lesssim 1600$ 
quasi-equilibrium HeII$\rightarrow$HeI recombination 
(according to the Saha formula) becomes valid.
\end{abstract}

\section{Introduction}
\hspace{0.9cm}
The HeII$\rightarrow$HeI primordial recombination has an 
effect on formation of the cosmic microwave 
background radiation (CMBR) anisotropy (e.g. Hu et al., 1995; 
Seager et al., 1999) 
and CMBR spectrum distortion (Lyubarsky and Sunyaev, 1983; Fahr and Loch, 1991; 
Dubrovich and Stolyarov, 1997; Wong et al., 2006).
The relative difference of CMBR anisotropy power spectrum corresponding 
to Saha HeII$\rightarrow$HeI recombination and corresponding to 
HeII$\rightarrow$HeI recombination by Seager et al. (1999) has a level 
of up to 5\% for multipoles $l\simeq 1500 - 3000$ (Seager et al., 2000). Such 
changes can be measured in future experiments (e.g. Planck and others). 

Kinetics of HeII$\rightarrow$HeI recombination has been considered in 
a number of papers (Boschan and Biltzinger, 1998; Novosyadlyj, 2006; 
Wong and Scott, 2006; and references therein). 
The most used model of helium recombination is the model suggested by 
Matsuda et al. (1969) and developed by Seager et al. (1999). 
A significant modification 
of this model was suggested by Dubrovich and Grachev (2005) 
who took into account recombination through HeI ortho-states. 

In the model by Seager et al. (1999) the effect of neutral hydrogen 
on HeII$\rightarrow$HeI recombination was neglected although 
in a number of previous papers (e.g. Hu et al., 1995; Boschan and 
Biltzinger, 1998; and others) authors pointed out that this effect 
may be essential. 
This contradiction made us investigate the effect of neutral hydrogen 
on HeII$\rightarrow$HeI recombination again and derive HeII$\rightarrow$HeI 
recombination kinetic equation which takes this effect into account.

\section{Addition to the recombination model}
The main addition to the Seager's et al. (1999) model is taking into account 
an interaction of resonant HeI quanta with neutral hydrogen. 
This interaction is described by the following 
elementary processes:
\\1. HeI resonant photon ionizes hydrogen atom with emission of 
electron which has a kinetic energy more than 6 eV: 
$H + \gamma \rightarrow H^+ + e^-$. 
\\2. The emitted electron loses the kinetic energy during 
electron-electron collisions with thermal ($E_{e,th}=3/2k_BT < 0.7 $ eV) 
electrons of the primordial plasma. 
The typical time of this process can be estimated by the following formula 
(e.g. Spitzer, 1978):
\begin{equation}
t_{ee}={m_{e}^2v_{e}^3 \over 4\pi N_{e} e^4
\ln\left(\Lambda m_{e}v_{e}^2 / 3k_BT\right)}\;,~~ 
\Lambda={3\over 2e^3}\sqrt{\left(k_BT\right)^3 \over \pi N_{e}}
\label{ee_time}
\end{equation}
where $m_{e}$ is the electron mass, $v_{e}$ is the electron velocity, 
$N_{e}$ is the free electron concentration, $T$ is the temperature of medium. 
Under considered conditions the value $t_{ee}$ is less than $10^3$ s.
\\3. The hydrogen ion recombines with a thermal electron and emits 
a photon with energy $\simeq (13.6+k_BT) \simeq 14$ eV: 
$H^+ + e^- \rightarrow H + \gamma$. 
The typical time of this process is larger than $10^9$ s.

Comparison of the typical times of electron-electron collisions and hydrogen 
recombination shows that between the ionization and 
the following recombination of hydrogen the energy of the non-thermal 
electrons is divided among the thermal electrons of the primordial plasma. 
 
In the chain of elementary processes mentioned above HeI resonant photons 
disappear and do not keep the populations of HeI excited states.

The rate of hydrogen ionization by HeI resonant $2p\rightarrow 1s$ photons 
$J_{H}$ [cm$^{-3}$s$^{-1}$] determines the rate of 
disappearance of these photons. 
It was shown by Seager et al. (2000) that this rate is 
negligible in comparison with the total rate of 
$2^1p\leftrightarrow 1^1s$ transitions (i.e. the 
difference of rates of forward $2^1p\rightarrow 1^1s$ and 
backward $1^1s\rightarrow 2^1p$ transitions) 
of HeI due to escape of resonant 
photons from line profile because of cosmological 
redshift $J_{HeI,bg}$ (subscript $b$ denotes $2^1p$ 
state of HeI, subscript $g$ denotes $1^1s$ state of HeI, fig. 1)
\footnote{Following Seager et al. (2000) we consider in this 
section only $2^1p\rightarrow 1^1s$ transitions, although similar 
consideration can be provided for HeI $2^3p\rightarrow 1^3s$ transitions.}.
In this work we have 
found that the rate $J_{H}$ was underestimated by Seager et al. (2000). 
To show this fact we considered the ratio of the rates 
$J_{H}$ and $J_{HeI,bg}$.

The rate $J_{H}$ is given by the expression
\begin{equation}
J_{H}\simeq c\sigma_{H}(\nu_{bg})N_{\gamma, bg}N_{HI}
\label{ion_rate1}
\end{equation}
where $c$ is the speed of light, $\sigma_H(\nu_{bg})$ is 
the hydrogen ionization cross section at frequency $\nu_{bg}$ of 
the $2^1p\rightarrow 1^1s$ transition of HeI, $N_{HI}$ is the concentration 
of hydrogen atoms in ground state, 
$N_{\gamma,bg}$ is HeI $2^1p\rightarrow 1^1s$ resonant photon 
concentration given by the formula
\begin{equation}
N_{\gamma,bg}\simeq{8\pi \nu_{bg}^3 \over c^3}\eta_{bg}
{\Delta \nu_D \over \nu_{bg}}=
{8\pi \nu_{bg}^3 \over c^3}\eta_{bg}
\sqrt{2k_BT \over m_{He}c^2}
\label{N_gamma_line}
\end{equation}
where $\eta_{bg}$ is the number of photons per mode 
$2^1p\rightarrow 1^1s$, $\Delta \nu_D$ 
is the thermal width of line $2^1p\rightarrow 1^1s$, 
$m_{He}$ is the mass of helium atom. 

The rate $J_{HeI,bg}$ [cm$^{-3}$s$^{-1}$] is given by the equation
\begin{equation}
J_{HeI,bg}\simeq A^{r}_{bg}N_{HeI,b}
\label{J_He2p1s}
\end{equation}
where $A^{r}_{bg}$ is the effective coefficient [s$^{-1}$] of 
$2^1p\leftrightarrow 1^1s$ transitions due to photon escape from the 
line profile 
because of cosmological redshift (superscript $r$ denotes ``redshifting''), 
$N_{HeI,b}$ is the concentration of HeI atoms in state $2^1p$. 
The effective coefficient of transitions is given by the formula:
\begin{equation}
A^{r}_{bg}\simeq{8\pi \nu_{bg}^3 H \over c^3 g_{b} N_{HeI,g}}
\label{A_red_HeI}
\end{equation}
where $H= H_0 \sqrt{\Omega_\Lambda+\Omega_m (1+z)^3 + \Omega_{rel}(1+z)^4}$ 
is the Hubble constant as function of redshift (parameters 
$H_0$, $\Omega_\Lambda$, $\Omega_m$ and $\Omega_{rel}$ described in tab. 1), 
$g_{b}=3$ is the statistical weight of $2^1p$ state, 
$N_{HeI,g}$ is the concentration of HeI atoms in $1^1s$ state.

Optical depth of absorption in the line HeI $2^1p\rightarrow 1^1s$ is much 
larger 
than unity, therefore $\eta_{bg}$ is satisfied the following expression 
\begin{equation}
\eta_{bg}\simeq{N_{HeI,b} \over g_{b} N_{HeI,g}}
\label{eta_21}
\end{equation}
Using equations (\ref{ion_rate1}) - (\ref{eta_21}) one can find the 
ratio of reaction rates
\begin{equation}
{J_{H} \over J_{HeI,bg}} \simeq 
{c\over H}\sigma_{H}(\nu_{bg}){\Delta \nu_D \over \nu_{bg}}N_{HI}
\label{JJ_ratio}
\end{equation}
In the case of $J_{H} / J_{HeI,bg} \ll 1$ the effect of neutral 
hydrogen on the HeII $\rightarrow$ HeI recombination is negligible. 
In the case of $J_{H} / J_{HeI,bg} \gg 1$ effect of neutral hydrogen 
dominates and recombination speeds up essentially.

Dependence of $J_{H} / J_{HeI,bg}$ (\ref{JJ_ratio}) on $z$ is shown in fig. 2. 
Exponential increase of $J_{H} / J_{HeI,bg}$ 
(with decrease of $z$) is concerned with exponential increase of 
neutral hydrogen concentration $N_{HI}$ which is satisfied the Saha 
ionization equation for HI in the epochs $z=1500 - 3000$.

The ratio $J_{H} / J_{HeI,bg}$ is equal to 1 at epoch 
$z\simeq 1800$ (fig. 2) in contradiction 
with results by Seager et al. (2000) where the ratio 
$J_{H} / J_{HeI,bg}$ is much less than unity in the period $z=1500-3000$.

Note that in paper by Boschan and Biltzinger (1998) other estimation 
of neutral hydrogen 
effect on helium recombination kinetics was proposed. In that work 
disappearance of HeI resonant photons was characterized by optical depth 
of hydrogen ionization by HeI $2^1p\rightarrow 1^1s$ resonant photons
\begin{equation}
\tau\simeq c\sigma_{H}(\nu_{bg})N_{HI}\Delta t 
\end{equation}
where $\Delta t$ is the duration of period when considered photon is the 
HeI resonant photon. The value $\Delta t$ is 
defined by deviation between photon frequency and central frequency of 
HeI resonant transition because only HeI resonant photons keep the 
populations of HeI excited states. Using $\dot \nu = -H\nu$ we find
\begin{equation}
\tau\simeq {c\over H}\sigma_{H}(\nu_{bg}){\Delta \nu_D \over \nu_{bg}}N_{HI}
\end{equation}

Thus one can obtain that
\begin{equation}
\tau\simeq {J_{H} / J_{HeI,bg}}
\label{theorem_1}
\end{equation}
and we can see that both of estimation (by reaction rate ratio and by 
optical depth) give the same results. The value $\tau$ is equal to 1 at 
$z \simeq 1800$ according to (\ref{theorem_1}) (see fig. 2). 
This result is similar to the result of paper by Boschan 
and Biltzinger (1998).

The simple estimation presented above shows that neutral hydrogen effect 
on helium recombination kinetics is essential. The fact made us consider 
this effect in detail and derive helium kinetic equation taking into account 
effect of neutral hydrogen.
\begin{figure}
\centering
\includegraphics[height=7cm, width=0.4\textwidth]{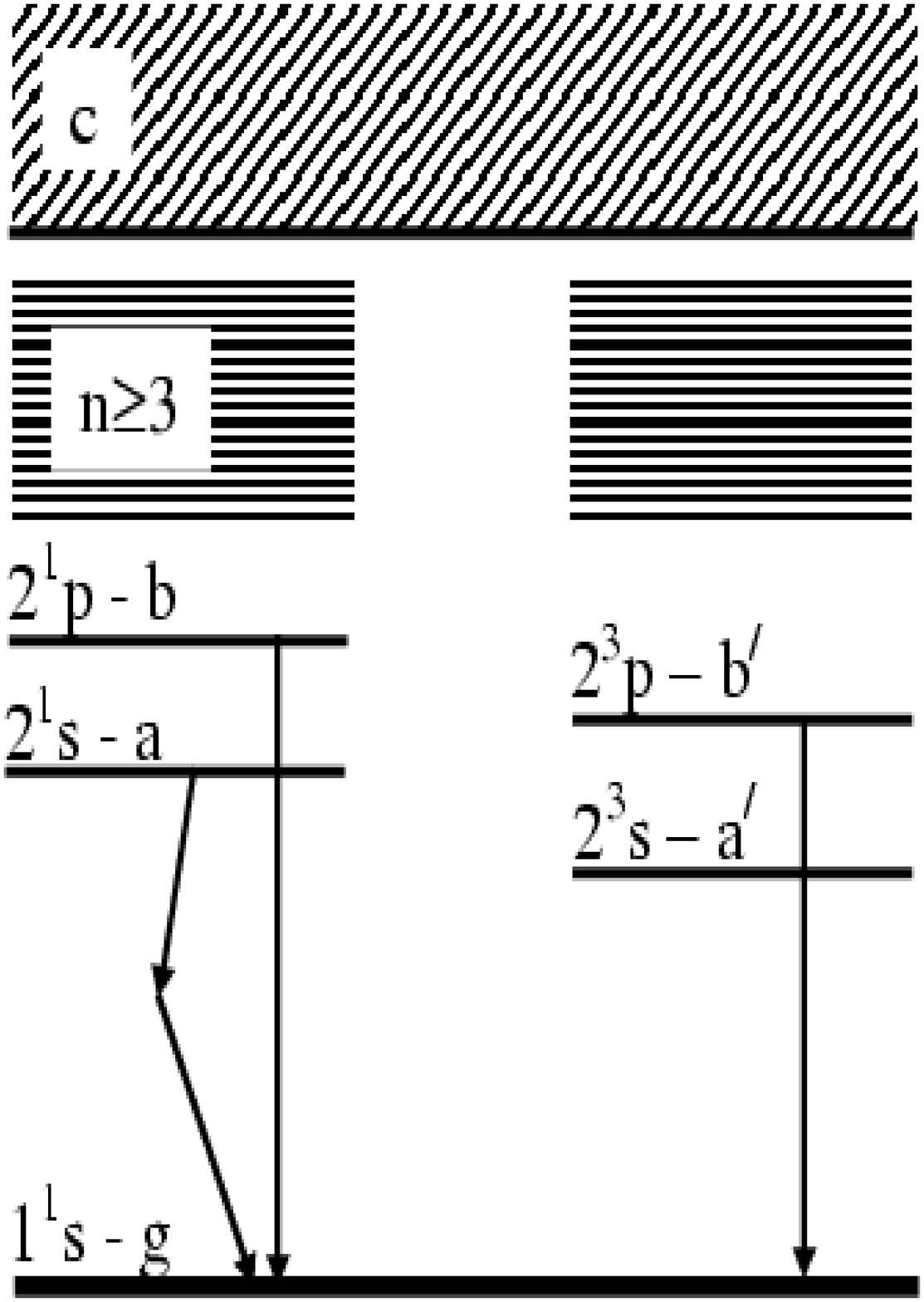}
\caption{HeI level structure used in this paper}
\label{level_scheme}
\end{figure}

\onecolumn
\section{Main equation}
The behaviour of HeII (i.e. He$^{+}$) fraction is described by the following 
differential equation
\begin{equation}
\dot x=-C_{par}\left(\alpha_{par}N_e x - 
{g_{a} \over g_{g}}\beta_{par}
\exp{\left(-{E_{ag}\over k_BT}\right)}(1-x) \right)
-C_{or}\left(\alpha_{or}N_e x - 
{g_{a'} \over g_{g} }\beta_{or}
\exp{\left(-{E_{a' g}\over k_BT}\right)}(1-x) \right)
\label{kin_eq_HeII}
\end{equation} 
where $x=N_{HeII}/N_{He}$ is the fraction of HeII ions 
relative to the total number of helium atoms and ions, 
$C_{par}$ is the factor by which the ordinary 
recombination rate is inhibited by the presence of HeI $2^1p\rightarrow 1^1s$ 
resonance-line radiation, $\alpha_{par}$ is the total HeII$\rightarrow$HeI 
recombination coefficient to the excited para-states of HeI, subscript $a$ 
denotes state $2^1s$ of HeI atom, $N_e$ is the free electron concentration, 
$g_{a}=1$ is the statistical weight of $2^1s$ state of HeI, $g_{g}=1$ is the 
statistical weight of $1^1s$ state of HeI, $\beta_{par}$ is the total 
HeI$\rightarrow$HeII ionization coefficient from the excited para-states of 
HeI, $E_{ag}$ is the $2^1s\rightarrow 1^1s$ transition energy, 
$C_{or}$ is the factor by which the ordinary 
recombination rate is inhibited by the presence of HeI $2^3p\rightarrow 1^1s$ 
resonance-line radiation, $\alpha_{or}$ is the total HeII$\rightarrow$HeI 
recombination coefficient to the excited ortho-states of HeI, 
subscript $a'$ denotes state $2^3s$ of HeI atom $g_{a'}=3$ is the 
statistical weight of $2^3s$ state of HeI, $\beta_{or}$ is the total 
HeI$\rightarrow$HeII ionization coefficient from the excited ortho-states of 
HeI, $E_{a'g}$ is the $2^3s\rightarrow 1^1s$ transition energy.

The para- and ortho- recombination and ionization coefficients are related 
by the following formulae
\begin{equation}
\beta_{par}={g_c \over g_{a}}\alpha_{par} g_e(T) 
\exp\left({-{E_{ca} \over k_B T}}\right), 
~~~~~~
\beta_{or}={g_c \over g_{a'}}\alpha_{or} g_e(T) 
\exp\left({-{E_{ca'} \over k_B T}}\right) 
\label{detailed_balance}
\end{equation}
where subscript $c$ denotes continuum state of HeI atom, 
$g_c=4$ is the statistical weight of continuum state of 
(He$^+$+e$^-$), $g_e(T)$ is the partition function of free electrons, 
$E_{ca}$ is the $c \rightarrow 2^1s$ transition energy, 
$E_{ca'}$ is the $c \rightarrow 2^3s$ transition energy. 

The inhibition factor $C_{par}$ is given by the following formula
\begin{equation}
C_{par}={\left(g_{b}/g_{a}\right)
\left(A^{H}_{bg}+A^{r}_{bg}\right)
\exp\left(-{E_{ba} / k_BT}\right)+A_{ag}
\over \beta_{par}+\left(g_{b}/g_{a}\right)
\left(A^{H}_{bg}+A^{r}_{bg}\right)
\exp\left(-{E_{ba} / k_BT}\right)+A_{ag}}
\end{equation}
where $A^{H}_{bg}$ is the effective coefficient [s$^{-1}$] of 
$2^1p\leftrightarrow 1^1s$ 
transitions due to effect of neutral hydrogen on HeI $2^1p\rightarrow 1^1s$ 
resonant radiation (superscript $H$ denotes hydrogen), 
$E_{ba}$ is the $2^1p\rightarrow 2^1s$ transition energy, 
$A_{ag}$ is the coefficient of two-photon $2^1s\rightarrow 1^1s$ spontaneous 
decay.

The inhibition factor $C_{or}$ is given by the following formula 
\begin{equation}
C_{or}={\left(g_{b'} / g_{a'} \right)
\left(A^{H}_{b'g}+A^{r}_{b'g}\right)
\exp\left(-{E_{b'a'}/ k_BT}\right) 
\over \beta_{or}+\left(g_{b'} / g_{a'} \right)
\left(A^{H}_{b'g}+A^{r}_{b'g}\right)
\exp\left(-{E_{b'a'}/ k_BT}\right)}
\end{equation}
where subscript $b'$ denotes state $2^3p$ of HeI, 
$g_{b'}=9$ is the statistical weight of $2^3p$ state of HeI, 
$A^{H}_{b'g}$ is the effective coefficient of $2^3p\leftrightarrow 1^1s$ 
transitions due to effect of neutral hydrogen on HeI $2^3p\rightarrow 1^1s$ 
resonant radiation, $A^{r}_{b'g}$ is the effective coefficient of 
$2^3p\leftrightarrow 1^1s$ 
transitions due to escape of photon from the $2^3p\rightarrow 1^1s$ line 
profile because of cosmological redshift, $E_{b'a'}$ is the 
$2^3p\rightarrow 2^3s$ transition energy.

The coefficient $A^{r}_{bg}$ is given by the formula (\ref{A_red_HeI}). 
The coefficient $A^{r}_{b'g}$ is given by the following formulae
\begin{equation}
A^{r}_{b'g}=A_{b'g}
\tau_{b'g}^{-1}\left(1-\exp\left(-\tau_{b'g}\right)\right),
~~~~~~~~~ \tau_{b'g}\simeq
{g_{b'} A_{b'g} c^3 N_{HeI,g} \over 8\pi H \nu_{b'g}^3}
\end{equation}
where $A_{b'g}=g_{2^3P_1}A_{2^3P_1\rightarrow 1^1S_0}/g_{b'}$ is the 
coefficient of $2^3p\rightarrow 1^1s$ spontaneous transition, 
$g_{2^3P_1}=3$ is the statistical weight of $2^3P_1$ state.

The coefficient $A^{H}_{fg}$ 
(subscript $f$ denotes $2^1p$ ($b$) or $2^3p$ ($b'$) state) can be obtained 
by consideration of kinetic equation for HeI $f\rightarrow g$ 
transitions and transfer equation for HeI $f\rightarrow g$ resonance radiation 
in the homogeneous expanding Universe in the presence of neutral hydrogen. 
Detailed calculation (paper in preparation) shows that exact expressions for 
the coefficients $A^{H}_{bg}$ and $A^{H}_{b'g}$ can be approximated by the 
following formulae:
\begin{equation}
A^{H}_{fg}=
{A_{fg} \over 1+p_{f}\gamma_{fg}^{q_{f}}},
~~~~~~~~~~~~
\gamma_{fg}=
{\left(g_{f}/g_{g}\right) A_{fg} N_{HeI} c^2 \over 
\sqrt{\pi} \sigma_{H}\left(\nu_{fg}\right) 8\pi
\nu_{fg}^2 \Delta \nu_{D,fg} N_{HI}}
\end{equation}
Parameters $p_{f}$ and $q_{f}$ are the following: $p_{b}=0.36$, 
$q_{b}=0.97$, $p_{b'}=0.66$, $q_{b'}=0.9$.  

The equation (\ref{kin_eq_HeII}) has been solved numerically. Results 
of calculation are presented in fig. 3.

The equation (\ref{kin_eq_HeII}) combined with equations for description 
of fractions of HII (Zel'dovich et al., 1968; Peebles, 1968) and HeIII 
(Matsuda et al., 1969; Seager et al., 1999) allows us to calculate 
ionization history of the Universe in the epochs $z=10^2 - 10^4$.
\twocolumn

\begin{figure}
\centering
\includegraphics[bb = 40 50 525 525, width=8cm]{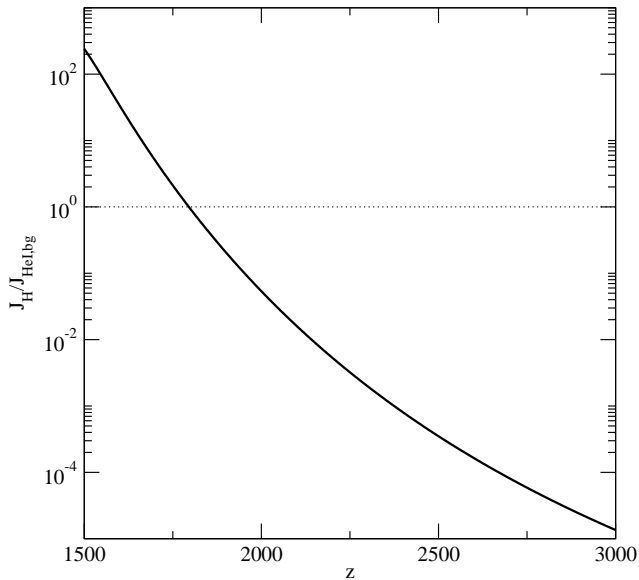}
\caption{The dependence of the ratio $J_{H} / J_{HeI,bg}$ 
on redshift $z$ is shown. The dotted curve shows the unity level.}
\label{fig1}
\end{figure}

\section{Radiation concerned with HeII$\rightarrow$HeI recombination}
The most of recombination photons corresponding to transitions 
to ground state of helium atoms is absorbed by the neutral hydrogen 
atoms and converts to the Ly$\alpha$-photons of HI. 

Number of HI Ly$\alpha$-photons per mode concerned with HeII$\rightarrow$HeI 
recombination is given by the formula
\begin{equation}
\Delta \eta (\nu_{\alpha}, z_b)=
{c^3 N_{He}\left(z_b\right) \over 8\pi \nu_{\alpha}^3}
{|\dot x(z_b)|\over H(z_b)}
\label{photon_number_1}
\end{equation}
where $\Delta \eta$ is the number of photons per mode depending on frequency 
$\nu$ and epoch $z$, $\nu_{\alpha}$ is the HI Ly$\alpha$ frequency, $z_b$ is 
the epochs of photon birth. Using relation
\begin{equation} 
\Delta \eta(\nu, z)=\Delta \eta\left(\nu {1+z' \over 1+z}, z'\right)
\end{equation}
we can find the spectrum of radiation concerned with HeII$\rightarrow$HeI 
recombination at the present epoch.

\begin{figure}
\centering
\includegraphics[bb = 25 55 540 740, width=8.4cm]{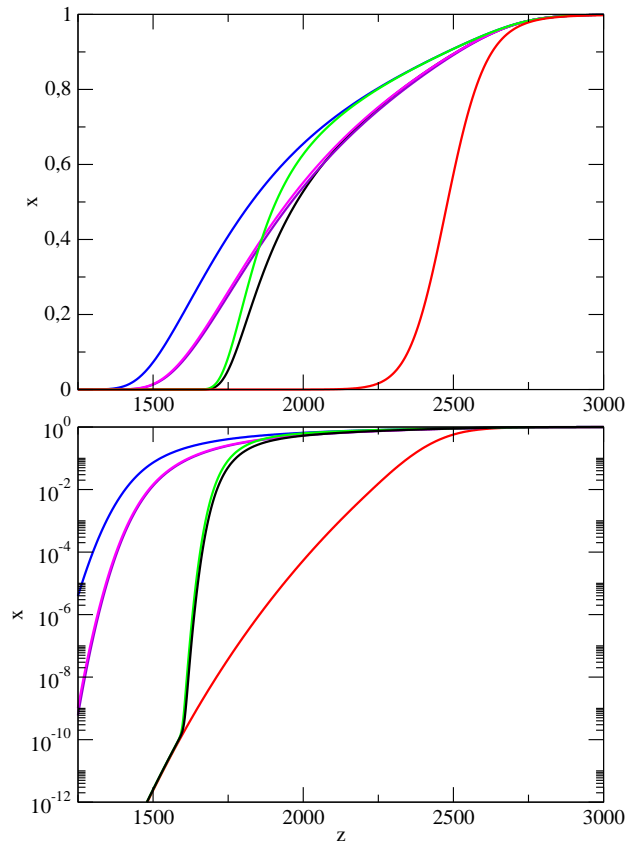}
\caption{The dependence of HeII fraction $x$ on redshift $z$ for 
various model of HeII$\rightarrow$HeI recombination is presented: 
blue curve corresponds to result of Seager et al., 1999, 
violet curve corresponds to taking into account recombination 
through ortho-states (Dubrovich and Grachev, 2005) 
and $A_{2^3P_1\rightarrow 1^1S_0}=233~s^{-1}$ (Lin et al., 1977), 
magenta curve corresponds to taking into account recombination 
through ortho-states (Wong and Scott, 2006) and 
$A_{2^3P_1\rightarrow 1^1S_0}=177.58~s^{-1}$ (Lach and Pachucki, 2001) 
[violet curve practically overlaps magenta curve], 
green curve corresponds to taking into account effect of neutral 
hydrogen on HeII$\rightarrow$HeI recombination, 
black curve corresponds to taking into account both of the effects, 
red curve corresponds to recombination according to the Saha formula for HeI. 
Top panel corresponds to linear scale, bottom one does logarithmic scale.}
\label{fig2}
\end{figure}

\begin{figure}
\includegraphics[bb = 35 55 525 525, width=8cm]{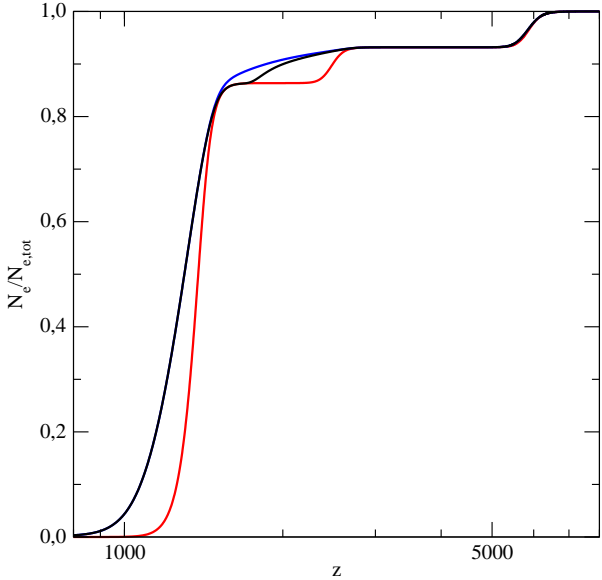}
\caption{The dependence of free electron fraction on redshift $z$ for 
various model of HeII$\rightarrow$HeI recombination is presented: 
blue curve corresponds to result of Seager et al., 1999, 
black curve corresponds this work, 
red curve corresponds to recombination according to the Saha formula. 
}
\label{fig4}
\end{figure}

\section{Results}
The main results of this paper are the dependencies of HeII fraction on 
redshift (fig. 3) and free electron fraction on redshift (fig. 4). 
The calculation shows that during epochs $z=1600-2000$ 
kinetics of HeII$\rightarrow$HeI recombination (this work, black curve) 
changes from strongly non-equilibrium (Seager et al., 1999, blue curve) 
to quasi-equilibrium (according to the Saha formula for HeI, red curve). 
This change is due to absorption of HeI resonant photons by neutral hydrogen 
whose concentration increases exponentially with decrease of temperature. 
At the epoch $z=1600$ the fraction of HeII relative to the total number of 
helium ions and atoms is less than $10^{-9}$. To compare our results 
with those of previous papers we have calculated 
HeII$\rightarrow$HeI recombination kinetics at the following parameters: 
$A^{H}_{fg}=0$, $A_{2^3P_1\rightarrow 1^1S_0}=233~s^{-1}$ 
(Lin et al., 1977) corresponding to paper by Dubrovich and Grachev (2005) - 
violet curve, and $A_{2^3P_1\rightarrow 1^1S_0}=177.58~s^{-1}$ 
(Lach and Pachucki, 2001) corresponding to paper by Wong and Scott (2006) - 
magenta curve. These results show very good agreement with results by 
Dubrovich and Grachev (2005) and Wong and Scott (2006). 

In the fig. 4 one can see that recombination of helium ends 
($x\le 10^{-6}$) before recombination of hydrogen begins 
($x_{HI}\ge 0.99$) in contradiction of results by Seager et al. (1999).
This change of primordial plasma recombination kinetics 
should led to changes of calculated CMBR anisotropy at the level 
of up to 5\% for multipoles $l \simeq 1500 - 3000$ (Seager et al., 2000).

Results of calculation of radiation intensity are presented in the fig. 5: 
1) Planck spectrum radiation at the temperature $T_0=2.726$ K (red curve); 
2) HI Ly$\alpha$ radiation concerned with hydrogen recombination (blue curve, 
Grachev and Dubrovich, 1991; Rubino-Martin et al., 2006; Kholupenko and
Ivanchik, 2006; and references therein);
3) HI Ly$\alpha$ radiation concerned with HeII$\rightarrow$HeI recombination
(magenta curve). 
Wien-tail CMBR intensity distortion concerned with HeII$\rightarrow$HeI 
recombination has the maximum value about $9.3\cdot 10^{-25}$ 
erg$\cdot$cm$^{-2}$ster$^{-1}$s$^{-1}$Hz$^{-1}$ being at frequency 1350 GHz. 
This frequency is about 2 times smaller 
than frequency calculated by Wong et al. (2006). This difference is 
concerned with re-emission of helium recombination photons by neutral 
hydrogen atoms in Ly$\alpha$-line of HI 
while in the paper by Wong et al. (2006) helium 
recombination photon emission was considered at transitions 
$2^1p\rightarrow 1^1s$ and $2^1s\rightarrow 1^1s$ of HeI.
\begin{figure}
\centering
\includegraphics[bb = 20 50 525 525, width=8cm, height=8cm]{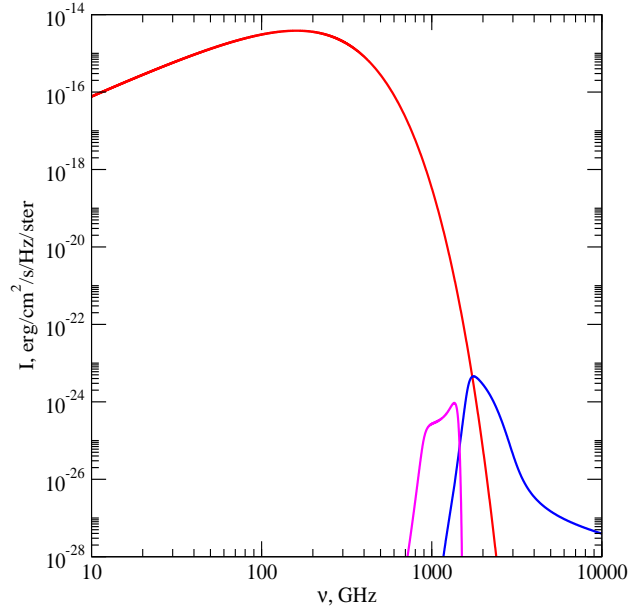}
\caption{The dependence of radiation intensity on frequency at present epoch: 
red curve corresponds to Planck radiation at temperature $T_0=2.726$ K, 
blue curve corresponds to HI Ly$\alpha$ radiation concerned with recombination 
of hydrogen, magenta curve corresponds to HI Ly$\alpha$ radiation concerned 
with recombination of helium (HeII$\rightarrow$HeI)}
\label{fig5}
\end{figure}
\vspace{1cm}
\begin{tabular}{lll}
  Table 1. Parameters of &the standard& cosmological\\model\\
  \hline
  Value description & Symbol & Value \\
  \hline
  total matter& $\Omega_{tot}$ & ~~~~1 \\
  non-relativistic matter& $\Omega_m$ & $0.27$ \\
  baryonic matter& $\Omega_b$ & $0.04$ \\
  relativistic matter& $\Omega_{rel}$ & $\sim 10^{-4}$ \\
  vacuum-like energy& $\Omega_\Lambda$ & $0.73$\\
  Hubble constant & $H_0$ & 70 km/s/Mpc \\
  radiation temperature & $T_{0}$ & $2.726$ K \\
  helium mass fraction & $Y$ & $0.24$ \\
  \hline
\end{tabular}
\label{lastpage}
\section*{Acknowledgments}
This work was supported by the Russian Foundation for Basic 
Research (project nos. 05-02-17065-a), the ``Leading Scientific 
Schools of Russia'' Program (NSh-9879.2006.2), and the Foundation 
for Support of Russian Science.

\end{document}